\documentclass[twocolumn]{aastex701}

\usepackage{amsmath,amssymb}
\usepackage{graphicx}
\usepackage{color}
\usepackage{soul}

\def\s{{\rm\thinspace s}}

\newcommand{\Msun}{M_{\odot}}

\def\paperI{\textbf{Paper~I}}
\def\coef{\rho_{\mathrm{Pearson}}}
\def\Lbol{L_{\mathrm{bol}}}
\def\MBH{M_{\mathrm{BH}}}
\def\REdd{\xi_\mathrm{Edd}}
\def\CGalDust{{r-\left<W1\right>}}
\def\Cmean{{\left<W1-W2\right>}}

\begin{document}

\title{A systematic study of the long-term mid-infrared color variations of Seyfert~1 galaxies}

\correspondingauthor{Liming Dou}
\email[show]{doulm@gzhu.edu.cn}

\author[0009-0003-1518-6186]{Jiahua Wu}
\affiliation{Department of Astronomy, Guangzhou University, Guangzhou 510006, China}
\email{jhwu@e.gzhu.edu.cn}

\author[0000-0002-4757-8622]{Liming Dou}
\affiliation{Department of Astronomy, Guangzhou University, Guangzhou 510006, China}
\email{doulm@gzhu.edu.cn}

\author[0009-0009-9334-1931]{Zhining Chen}
\affiliation{Department of Astronomy, Guangzhou University, Guangzhou 510006, China}
\email{czning@e.gzhu.edu.cn}

\author[0000-0002-0403-9522]{Fupeng Zhang}
\affiliation{Department of Astronomy, Guangzhou University, Guangzhou 510006, China}
\email{zhangfupeng@gzhu.edu.cn}

\author[0000-0001-6947-5846]{Luis C. Ho}
\affiliation{Kavli Institute for Astronomy and Astrophysics, Peking University, Beijing 100871, China}
\affiliation{Department of Astronomy, School of Physics, Peking University, Beijing 100871, China}
\email{lho.pku@gmail.com}

\author[0000-0003-4897-4106]{Yanli Ai}
\affiliation{College of Engineering Physics, Shenzhen Technology University, Shenzhen 518118, People’s Republic of China}
\email{aiyanli@sztu.edu.cn}

\author[0000-0002-1517-6792]{Tinggui Wang}
\affiliation{Department of Astronomy, University of Science and Technology of China, Hefei, 230026, China}
\email{twang@ustc.edu.cn}
 
\author[0000-0002-7020-4290]{Xinwen Shu}
\affiliation{Department of Physics, Anhui Normal University, Wuhu, Anhui 241002, China}
\email{xwshu@ahnu.edu.cn}

\author[0000-0002-7152-3621]{Ning Jiang}
\affiliation{Department of Astronomy, University of Science and Technology of China, Hefei, 230026, China}
\email{jnac@ustc.edu.cn}
 
\author[0000-0003-4874-0369]{Junfeng Wang}
\affiliation{Department of Astronomy, Xiamen University, Xiamen, Fujian 361005, China}
\email{jfwang@xmu.edu.cn}

\begin{abstract}
We present a comparative analysis of long-term mid-infrared (MIR) color variations (MCVs) for 1,977 narrow-line Seyfert~1 (NLSy1) and 4,113 broad-line Seyfert~1 (BLSy1) galaxies at $z < 0.3$. Using 14 years of WISE and NEOWISE data, we quantify the correlation between $W1-W2$ color and $W1$ magnitude. We categorize sources into bluer-when-brighter (BWB), redder-when-brighter (RWB), and weak/no MCV populations. Our results show that the MCV properties of BLSy1s are consistent with those of NLSy1s. This suggests that host-contaminated dust reprocessing is a universal mechanism across Seyfert~1s. Bolometric luminosity ($\Lbol$) is the primary driver of MCV behavior. However, at fixed $\Lbol$, the fraction of RWB sources decreases with increasing Eddington ratio ($\REdd$). This reveals a secondary dependence on the accretion state. Ensemble structure function analysis further shows that higher-$\REdd$ sources exhibit flatter structure functions at a given $\Lbol$, indicating smaller dust torus radii. These findings suggest that the central engine's radiation and accretion state shape the circumnuclear dust geometry. Radio-loud sources follow the same trends as radio-quiet counterparts, implying that jet emission does not significantly impact long-term MIR variability. Finally, modified blackbody modeling indicates that grains with sizes $a \gtrsim 0.1~\mu$m are responsible for the observed MCVs. Our results establish a unified framework for MCVs in Seyfert~1s, where the Eddington ratio modulates both variability patterns and the physical scale of the dust torus. Furthermore, given the similarity between the MCV distributions of CLAGNs and RWB sources, we estimate that $\lesssim 6\%$ of Seyfert~1s may be CLAGN candidates, offering a potential MIR-based pre-selection.
\end{abstract}


\section{Introduction}\label{sec:intro}
Active galactic nuclei (AGN; \citealt{Seyfert1943, Schmidt1963}) are powered by mass accretion onto supermassive black holes (SMBHs; $10^6$--$10^{10}~\Msun$) at the centers of galaxies.
Their radiative output is closely coupled to the physical conditions of the accretion flow and the circumnuclear material. The mid-infrared (MIR) emission of AGN is dominated by thermal radiation from circumnuclear dust heated by optical/UV photons from the accretion disk \citep{Barvainis1987}, providing a direct view of the dust distribution around the central engine.
Such circumnuclear dust is generally thought to be distributed in a toroidal structure that can obscure the innermost regions along certain lines of sight \citep{Antonucci1993}.
Seyfert~1 galaxies are particularly valuable laboratories in this context, as they provide an unobscured view of both the central engine and the innermost, hottest dust grains.
These sources are commonly divided into narrow-line Seyfert~1 (NLSy1) and broad-line Seyfert~1 (BLSy1) subclasses according to the widths of their permitted emission lines \citep{Osterbrock_and_Pogge1985, Goodrich1989}. Compared with BLSy1s, NLSy1s are generally associated with lower-mass SMBHs ($\MBH \sim 10^{6}$--$10^{7}~\Msun$) and higher Eddington ratios ($\REdd$), whereas BLSy1s typically host more massive SMBHs ($\MBH \sim 10^{8}$--$10^{9}~\Msun$) and accrete at lower $\REdd$ \citep[e.g.,][]{Peterson2000, Grupe2010}.

Recent observational evidence suggests that the torus structure is dynamically regulated by the accretion state of the central engine. Reverberation mapping (RM) techniques have directly probed the torus radius by measuring the time delay between variations in the optical/UV continuum and the subsequent response in the MIR. These studies have robustly established a relationship between the torus radius ($R$) and the AGN bolometric luminosity ($\Lbol$), following the form $\log R \propto 0.5 \log \Lbol$ \citep[e.g.,][]{Koshida2014, Lyu2019, Yang2020, Tomar2025}. This scaling relation, further corroborated by \textit{GRAVITY} interferometric measurements \citep{GRAVITY2023}, is physically interpreted as the dust sublimation radius defined by the incident radiation field.

Beyond the radial scale, the torus covering factor --- the fraction of central emission intercepted and reprocessed by dust --- appears to be sensitive to the accretion rate. Utilizing a systematic multi-wavelength survey of hard-X-ray-selected AGN, \citet{Ricci2017} demonstrated that radiation pressure on dusty gas is a primary mechanism for regulating the distribution of circumnuclear material. Specifically, they observed that the covering factor decreases as the mass-normalized accretion rate increases, a trend supported by subsequent investigations \citep[e.g.,][]{Ricci2023, Mizukoshi2024}. This reduction is likely attributable to radiation-driven outflows that clear the circumnuclear dusty gas at high Eddington ratios.

Further complexities in this geometry were highlighted by \citet{Zhuang2018}, who performed detailed infrared spectral energy distribution (SED) modeling of Palomar-Green quasars. They found that the torus opening angle declines with increasing $\REdd$ up to $\sim 0.5$, beyond which the trend reverses and the opening angle widens. This reversal was attributed to a transition in the accretion flow geometry, shifting from a standard geometrically thin disk to a geometrically thick slim disk as accretion rates approach the super-Eddington regime. Collectively, these findings underscore that the geometry and physical state of the dust torus are intrinsically coupled to the central accretion conditions.

If the accretion state modulates the torus structure, observable signatures should manifest in the variability properties of Seyfert~1 galaxies. Recent investigations of the MIR ensemble structure function (SF) provide an independent and critical perspective on this link. The MIR SF characterizes how variability amplitude scales with rest-frame time lag ($\Delta t$); specifically, its slope has been shown to correlate with the inner radius ($R_{\rm in}$) of the dust torus, where a flatter slope indicates a smaller inner radius \citep{Li_and_Shen2023}. Furthermore, \citet{Hu_and_Mao2025} found that NLSy1s exhibit systematically flatter MIR ensemble SFs compared to BLSy1s, suggesting intrinsic differences in torus geometry that may be fundamentally linked to their distinct accretion parameters.

While the SF captures statistical variability amplitudes, the MIR color variation (MCV) offers a complementary and more physically direct window into the evolution of dust temperature. The development of the \textit{Wide-field Infrared Survey Explorer} \citep[WISE;][]{Wright2010} has enabled decade-long, all-sky MIR monitoring, revealing that AGN generally exhibit larger MIR variability amplitudes than inactive galaxies due to the thermal response of circumnuclear dust. Although the MIR color criterion $W1 - W2 \geq 0.8$ effectively isolates AGN from stellar and galactic populations \citep{Assef2013}, its temporal behavior remains less systematically explored.
Subsequent studies have shown that the MIR colors of AGN can evolve in opposite directions as source brightness varies. A redder-when-brighter (RWB) trend is frequently reported and often interpreted as a signature of an increased dust contribution relative to the invariant stellar continuum of the host galaxy \citep{Son2022}. Conversely, bluer-when-brighter (BWB) behavior is observed when the hot dust SED peak shifts blueward upon enhanced heating \citep{jhwu2026}, or when non-thermal jet emission introduces additional complexity \citep{Anjum2020}. 

In our previous systematic study of 1,718 low-redshift ($z < 0.3$) NLSy1s \citep[][hereafter \paperI{}]{jhwu2026}, we utilized 14 years of WISE and NEOWISE data to establish a foundational framework for long-term MCV. We identified RWB trends in 133 sources and BWB trends in 235 sources, with the majority of the sample showing weak/no MCV. These patterns were interpreted within a physical framework of host-contaminated dust reprocessing, where the thermal response of the dust to variable accretion luminosity is diluted by the stable host-galaxy starlight. Critically, \paperI{} identified bolometric luminosity as the primary driver of MCV, showing that high-$\Lbol$ sources preferentially exhibit stronger BWB trends, seemingly independent of black hole mass or Eddington ratio. Furthermore, we found that the MCV properties of radio-loud (RL) NLSy1s did not differ significantly from those of their radio-quiet (RQ) counterparts, despite the potential for jet contamination.

These findings raise important questions regarding the universality of MCV phenomena. It remains unclear whether the observed long-term MCV features are unique to the physical conditions of NLSy1s, such as their high accretion rates and relatively small black hole masses, or represent a ubiquitous feature of the AGN dust torus. BLSy1s, which typically host more massive SMBHs and/or lower $\REdd$, provide a crucial counterpoint. A systematic comparison between these two subclasses is necessary to determine whether MCV behavior is primarily driven by central engine luminosity or by other fundamental AGN parameters. By investigating the broader Seyfert~1 population, we can test whether the $\Lbol$-driven interpretation proposed in our previous work holds across a wider range of physical states, or if factors such as $\MBH$ and $\REdd$ play a more significant role than previously recognized.

This paper is organized as follows: Section~\ref{sec:sample_and_data} describes the sample selection and WISE data processing. Section~\ref{sec:quantification} details our MCV classification methodology. Section~\ref{sec:mcv_correlation} and Section~\ref{sec:mcv_fraction} present the statistical comparison between NLSy1s and BLSy1s. The ensemble structure functions are analyzed in Section~\ref{sec:esf}. We discuss physical implications in Section~\ref{sec:discussion} and summarize our findings in Section~\ref{sec:conclusions}.
Throughout this work, we assume a flat $\Lambda$CDM cosmology with $H_0 = 70\,\text{km s}^{-1}\,\text{Mpc}^{-1}$, $\Omega_{\text{M}} = 0.3$, and $\Omega_{\Lambda} = 0.7$. Unless otherwise specified or indicated, all magnitudes are reported in the Vega system.

\section{Sample Selection and Data Reduction\label{sec:sample_and_data}}
\subsection{Seyfert~1 Galaxies Catalog}

The initial sample for this study is drawn from the comprehensive Seyfert~1 galaxy catalog of \citet{Paliya2024}, which contains 52,273 broad-line Seyfert~1 (BLSy1) and 22,656 narrow-line Seyfert~1 (NLSy1) galaxies. This parent catalog was derived by applying the \textit{Bayesian AGN Decomposition Analysis for SDSS Spectra} (\texttt{BADASS}, \citealt{Sexton2021}) to optical spectra from the Sloan Digital Sky Survey Data Release 17 (SDSS-DR17). To minimize potential biases from cosmological evolution and to reduce uncertainties associated with $K$-corrections, we restrict our analysis to a low-redshift subsample with $z < 0.3$.

\subsection{WISE Multi-epoch Photometric Data}

To characterize long-term MCVs, we cross-matched the optical coordinates of the $z < 0.3$ Seyfert~1 sample with the WISE database. The WISE scanning strategy, which covers the entire sky approximately every six months, provides a naturally binned time series ideal for probing the dust torus through MIR variability.

We utilize data from both the AllWISE and NEOWISE Reactivation mission phases \citep{Wright2010, Mainzer2014}. Our analysis is restricted to the $W1$ (3.4~$\mu$m) and $W2$ (4.6~$\mu$m) bands, as the $W3$ and $W4$ bands are unavailable during the NEOWISE-R phase. We performed a cross-match using a 6\arcsec\ search radius centered on the SDSS positions. To ensure high data fidelity, the following quality criteria were applied to the single-exposure photometry:
\begin{enumerate}
    \item Only framesets with \texttt{qual\_frame} $> 0$ were retained to exclude spurious detections or transient artifacts.
    \item We required a reduced $\chi^2 < 5$ for the profile-fit photometry in both $W1$ and $W2$ (i.e., \texttt{w1rchi2} $< 5$ and \texttt{w2rchi2} $< 5$).
    \item To minimize blending and host galaxy confusion, the number of point-spread function (PSF) components used in the profile fit was limited to \texttt{nb} $\leq 2$.
    \item We retained only high-quality exposures unaffected by known artifacts (\texttt{cc\_flags} = `00'), moon contamination (\texttt{moon\_masked} = 0), or active de-blending (\texttt{na} = 0).
    \item Framesets obtained during South Atlantic Anomaly (SAA) passage (\texttt{saa\_sep} $\leq 0$) were excluded.
\end{enumerate}

\subsection{Spatial Clustering via DBSCAN}
While the initial 6\arcsec\ match ensures completeness, it introduces potential contamination from nearby sources. We employed the \textit{Density-Based Spatial Clustering of Applications with Noise} (DBSCAN) algorithm \citep{Ester1996} to isolate data points belonging to the primary AGN. The DBSCAN parameters were set to a maximum neighborhood distance of 0\farcs5 and a minimum of five points to define a cluster core. We retained the cluster whose centroid was closest to the optical SDSS position.

\begin{figure}[hbt!]
\centering
\includegraphics[width=\columnwidth]{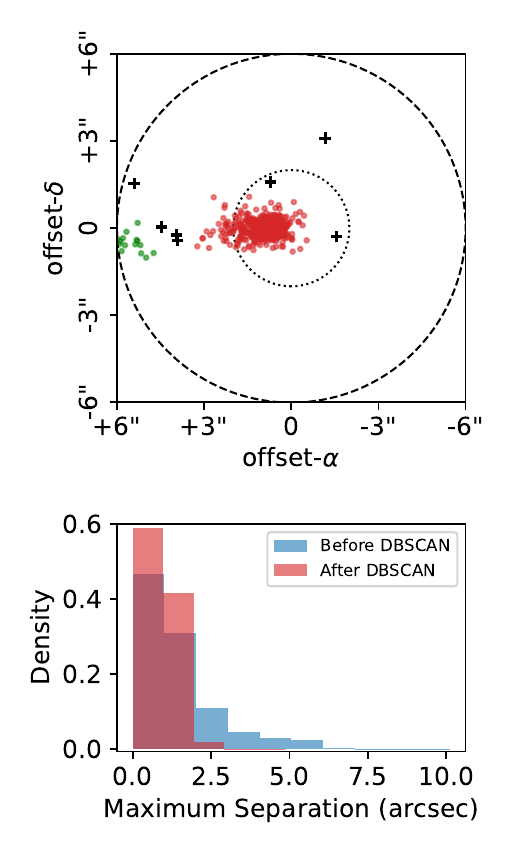}
\caption{Spatial clustering and cleaning of WISE photometry. 
Upper: Illustrative example of the DBSCAN algorithm applied to the NLSy1 source SDSS~J120847.81+671833.8 ($z=0.249$). The axes represent positional offsets relative to the SDSS optical coordinates. The inner and outer dashed circles indicate search radii of 2\arcsec\ and 6\arcsec, respectively. Red circles identify detections associated with the target AGN, while green circles denote classified contaminants (from galaxy SDSS~J120849.12+671832.7). Black plus signs represent data points identified as noise by the algorithm. 
Lower: Distributions of the maximum angular separation between individual WISE photometric detections and the median WISE coordinate for the entire sample. The blue histogram represents the distribution following initial cross-matching and quality filtering, while the red histogram shows the significant reduction in positional scatter after applying the DBSCAN cleaning procedure.
\label{fig:dbscan}
}
\end{figure}

This procedure effectively filtered out non-associated contaminants (see Figure~\ref{fig:dbscan}, top panel). Post-cleaning, 98.4\% of the sources exhibited positional scatter within 2\arcsec\ of the median WISE coordinates. Figure~\ref{fig:dbscan} (bottom panel) illustrates the significant reduction in angular separation achieved by the DBSCAN cleaning compared to the raw cross-match.

\subsection{Outlier Rejection and Light-curve Construction}
To mitigate the impact of cosmic rays and other transient outliers, we applied a 5$\sigma$ clipping to the single-exposure measurements within each approximately six-month observational epoch. 

For the final long-term light curves, we calculated epoch-binned magnitudes. We required a minimum of five single-exposure detections per epoch and at least five distinct epochs per source to ensure sufficient temporal coverage. The epoch-averaged magnitude ($m_{\text{epoch}}$) was determined by converting single-exposure Vega magnitudes ($m_i$) to flux densities, averaging the flux, and converting back to magnitudes. We adopted zero-point fluxes of $F_{\nu,0} = 309.540$~Jy for $W1$ and 171.787~Jy for $W2$ \citep{Wright2010}. The associated uncertainty ($\varepsilon_e$) for each epoch was calculated as:
\begin{equation}
\varepsilon_e^2 = \frac{1}{N_s(N_s-1)}\sum_{i=1}^{N_s} (m_i - m_{\text{epoch}})^2 + \frac{1}{N_s^2}\sum_{i=1}^{N_s}\varepsilon^2_i + \frac{1}{N_s}\varepsilon^2_{\mathrm{sys}},
\end{equation}
where $N_s$ is the number of exposures per epoch, $\varepsilon_i$ is the individual measurement error, and $\varepsilon_{\mathrm{sys}} = 0.016$~mag is the systematic uncertainty adopted from \citet{Lyu2019}.

\section{Quantification and Classification of Mid-infrared Color Variation}\label{sec:quantification}
We constructed an MIR color curve, defined as $C = W1 - W2$, for each source in our sample.
We quantified the MCV behavior by calculating the Pearson correlation coefficient ($\coef$) between the color $C$ and the $W1$ magnitude. Under this definition, a source exhibits a bluer-when-brighter (BWB) trend if $\coef > 0$ and a redder-when-brighter (RWB) trend if $\coef < 0$, where values approaching $\pm 1$ indicate strong monotonic relationships.

\subsection{Variability Significance}

To ensure the reliability of the derived $\coef$, we excluded sources with low-significance MIR variability. We estimated the intrinsic variability amplitude, $\sigma_m$, following the formalism of \citet{Sesar2007}. The observed scatter in the epoch-binned magnitudes is given by
\begin{equation}
\Sigma = \sqrt{ \frac{1}{N-1} \sum_{i=1}^{N} (m_{\text{epoch},i} - \overline{m}_{\text{epoch}})^2 },
\end{equation}
where $N$ is the number of epochs, $m_{\text{epoch},i}$ is the magnitude of the $i$-th epoch, and $\overline{m}_{\text{epoch}}$ is the mean magnitude across all epochs. The root-mean-square of the epoch uncertainties is defined as
\begin{equation}
\varepsilon_{\text{rms}} = \sqrt{\frac{1}{N}\sum_{i=1}^{N} \varepsilon_{e,i}^2 },
\end{equation}
where $\varepsilon_{e,i}$ is the uncertainty of $m_{\text{epoch},i}$. The intrinsic scatter, corrected for measurement noise, is then
\begin{equation}
\sigma_m =
\begin{cases}
\sqrt{ \Sigma^2 - \varepsilon^2_{\text{rms}} }, & \text{if } \Sigma > \varepsilon_{\text{rms}}, \\
0, & \text{otherwise}.
\end{cases}
\end{equation}
We retained only those sources satisfying the condition $\sigma_m > 2\varepsilon_{\text{rms}}$ in both the $W1$ and $W2$ bands.

\subsection{Monte Carlo Uncertainty and Significance Estimation}
\label{sec:mc_estimation}
We employed a Monte Carlo perturbation approach to account for the impact of photometric uncertainties on $\coef$. We first estimated the resulting uncertainty in $\coef$. For each source, we generated 1,000 realizations of the $W1$ and $W2$ light curves by perturbing each epoch's magnitude with a random deviation drawn from a Gaussian distribution, $\mathcal{N}(0, \varepsilon^2_e)$. We computed $\coef$ for each realization to determine its distribution. The $1\sigma$ uncertainty of $\coef$ was defined as the half-width of the interval between the 16th and 84th percentiles. For our sample, the median $1\sigma$ uncertainty is 0.12.

In \paperI{}, we adopted $\left|\coef\right|>0.6$ as the criterion for identifying significant MCV. To assess the statistical significance of this threshold, we performed an additional Monte Carlo test for the entire sample under the null hypothesis that there is no correlation between the $W1$ magnitude and the color $C$. In each realization, we randomly selected a source from our sample and perturbed its $W1$ and $W2$ magnitudes following the uncertainty estimation procedure described above. We then randomly permuted the temporal order of the color measurements to remove any correlation between $W1$ and $C$ while preserving their individual distributions, and recalculated $\coef$. Repeating this procedure $N_{\rm MC}=10^5$ times yielded a sample-wide null distribution that accounts for the observed sampling and photometric uncertainties of the entire sample.

The resulting null distribution is well centered around zero ($=-0.0002$), indicating that the permutation procedure effectively removes any correlation between $W1$ and $C$. From the resulting null distribution, we estimated the probability of obtaining an absolute MCV coefficient larger than a given threshold $r_0$ as
\begin{equation}
P_{\rm null}(r_0) = \frac{
N\left(\left|\coef\right|>r_0\right)
}{N_{\rm MC}}.
\end{equation}
We find that the commonly adopted significance levels of $P_{\rm null}=0.05$ and $0.001$ correspond to $r_0=0.43$ and $0.76$, respectively. For the empirical threshold of $r_0=0.6$ adopted in \paperI{}, we obtain $P_{\rm null}=6.84\times10^{-3}$. We therefore retain $\left|\coef\right|>0.6$ for consistency with \paperI{} and to facilitate direct comparison with its results.

\subsection{Classification of MCV Trends}
Our final sample consists of 1,977 NLSy1s and 4,113 BLSy1s that meet all redshift, data-quality, and variability-significance criteria. Sources were further classified by radio-loudness ($R \equiv f_{\text{5\,GHz}}/f_{4400\,\text{\AA}}$), with $R < 10$ defining radio-quiet (RQ) and $R \geq 10$ defining radio-loud (RL) populations \citep{Paliya2024}. The 5~GHz flux was extrapolated from 1.4~GHz FIRST survey data \citep{White1997} assuming a spectral index of $\alpha=0.5$.

\begin{figure*}[hbt!]
\centering
\plotone{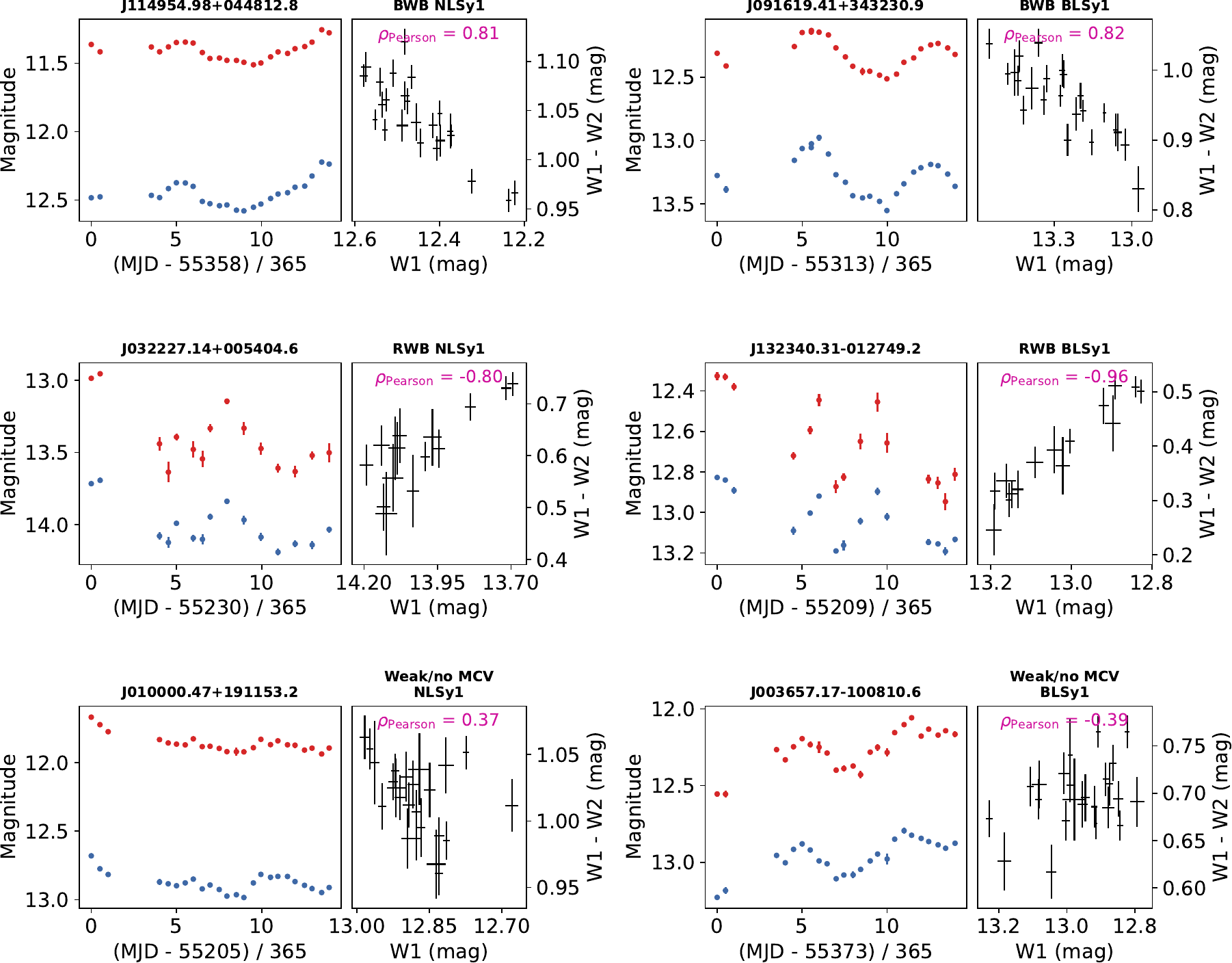}
\caption{
Example MIR light curves for each classification in Section~\ref{sec:quantification}. For each panel, light curves (left; blue in the $W1$ band and red in the $W2$ band) and color variations following the photometric magnitudes in the $W1$ band (right) are shown.
\label{fig:lc_demos}}
\end{figure*}

Based on $\coef$ and the associated null-hypothesis probability ($P_{\text{null}}$), we categorize the sources into three groups:
\begin{enumerate}
    \item \textit{BWB Sy1s:} 989 sources with a strong positive correlation ($\coef > 0.6$ and $P_{\text{null}} < 0.05$).
    \item \textit{RWB Sy1s:} 379 sources with a strong negative correlation ($\coef < -0.6$ and $P_{\text{null}} < 0.05$).
    \item \textit{Weak/no MCV Sy1s:} 4,722 sources with weak ($|\coef| \leq 0.6$) or statistically insignificant ($P_{\text{null}} \geq 0.05$) color variation.
\end{enumerate}
The census of these subsamples is summarized in Table~\ref{tab:number_mcv}. Representative light curves and color-magnitude diagrams for each category are displayed in Figure~\ref{fig:lc_demos}.

\section{Properties of Seyfert 1 galaxies with Mid-infrared Color Variation}\label{sec:mcv_correlation}

To examine whether the MCV behaviors identified in NLSy1s extend to the broader Seyfert 1 population, we applied the analytical framework established for NLSy1s in \paperI{} to our combined Seyfert~1 sample.\footnote{While the parent samples here and \paperI{} both originate from the catalog of \citet{Paliya2024}, the NLSy1 subsample in this study is defined after applying more stringent data-reprocessing steps as described in Section~\ref{sec:sample_and_data} and \ref{sec:quantification}. The resultant statistical conclusions regarding NLSy1 MCVs remain fully consistent with the earlier work.} This comparative analysis focuses on correlations between $\coef$ and key AGN properties, including variability amplitude (Section~\ref{sec:sigma_m}), MIR colors (Section~\ref{sec:colors}), and fundamental black hole parameters ($\MBH$, $\Lbol$, and $\REdd$; Section~\ref{sec:agn_params}). Pearson correlation coefficients and linear regression fits for these relationships are summarized in Table~\ref{tab:coef}. Finally, we analyze the distribution of NLSy1s and BLSy1s in the $C_{\rm min}$--$\Delta C$ plane (Section~\ref{sec:cplane}).

\subsection{Intrinsic variability amplitude}\label{sec:sigma_m}
\begin{figure*}[hbt!]
\centering
\plotone{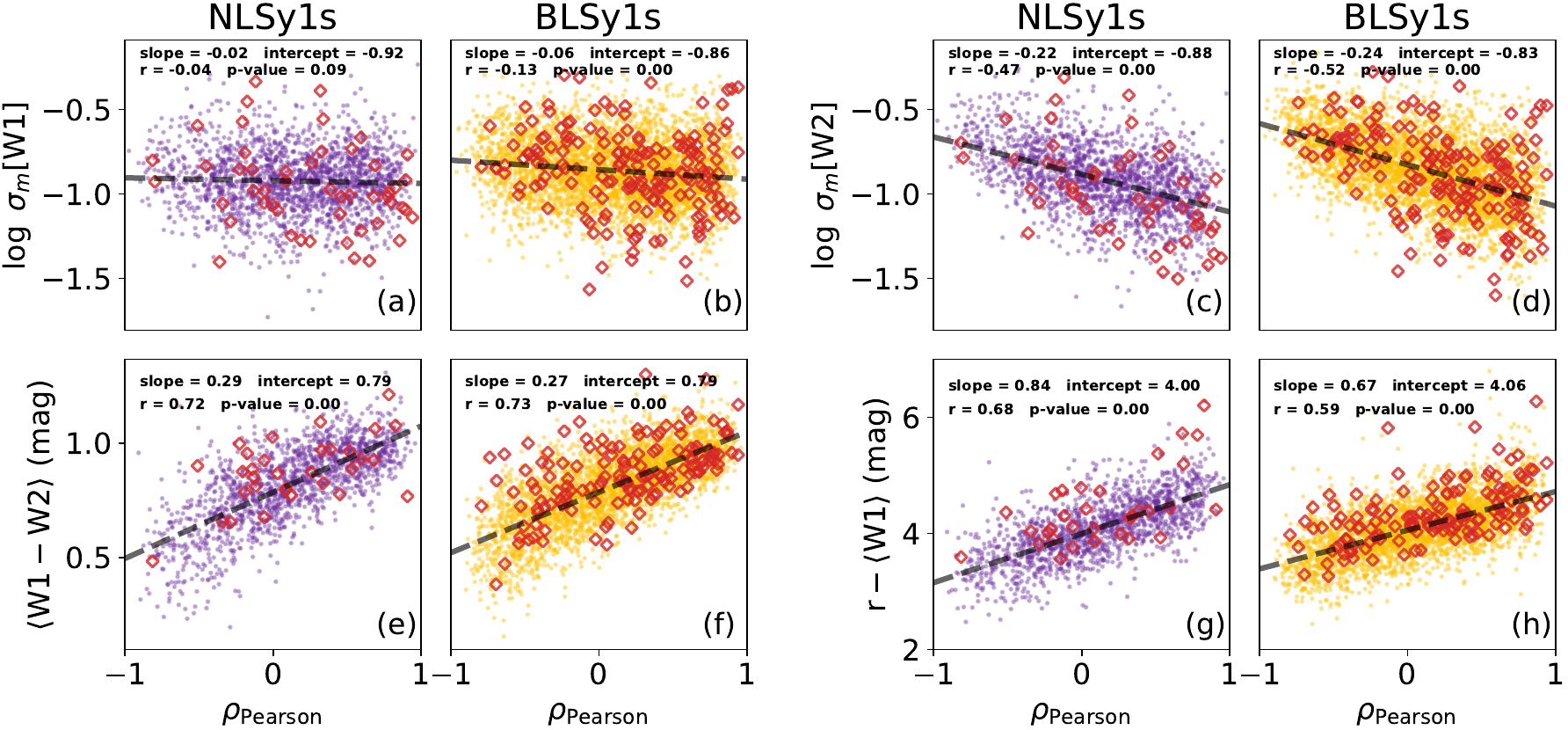}
\caption{
Correlations between the MCV coefficient $\coef$ and MIR variability properties.
(a--d) $\coef$ versus intrinsic variability amplitude $\sigma_m$ in the $W1$ and $W2$ bands for NLSy1s and BLSy1s.
(e--f) $\coef$ versus mean MIR color $\Cmean$.
(g--h) $\coef$ versus the host-contamination proxy $\CGalDust$.
Purple and orange points denote RQ-NLSy1s and RQ-BLSy1s, respectively; red diamonds indicate radio-loud counterparts. Dashed lines represent linear regression fits for the RQ populations.
\label{fig:mcv_analy}}
\end{figure*}

We first investigated the relationship between the MIR intrinsic variability amplitude $\sigma_m$ and the color variation coefficient $\coef$. For RQ-BLSy1s, the Pearson correlation coefficients between $\log \sigma_m$ and $\coef$ are $-0.13$ (p-value $\ll 0.001$) and $-0.52$ (p-value $\ll 0.001$) for the $W1$ and $W2$ bands, respectively. These results are consistent with those reported for RQ-NLSy1s (-0.04 in $W1$ and -0.47 in $W2$; \paperI{}), as visualized in Figure~\ref{fig:mcv_analy}.

For both subclasses, the correlation is significantly stronger in the $W2$ band than in $W1$. Linear regression shows no significant difference in the slope of $\log \sigma_m$ versus $\coef$ between NLSy1s and BLSy1s. The slope is approximately $0$ in $W1$ and $-0.23$ in $W2$ (see Table~\ref{tab:coef} for details). This similarity suggests a common mechanism governing the relationship between variability amplitude and color trend in both populations. Radio-loud (RL) sources (red diamonds in Figure~\ref{fig:mcv_analy}) closely follow the distribution of their RQ counterparts, consistent with trends observed for RL-NLSy1s.

\subsection{MIR Colors and Host Contamination}\label{sec:colors}

The mean MIR color, $\Cmean$, serves as a proxy for the characteristic dust temperature, while $\CGalDust$ traces the relative contribution of the host galaxy to the MIR flux. Both Seyfert~1 subclasses exhibit positive correlations between $\coef$ and these colors (Figure~\ref{fig:mcv_analy}). The correlation coefficient for $\Cmean$ is $\sim 0.72$, with a slope of $\sim 0.28$. For $\CGalDust$, the correlation coefficients are 0.68 (RQ-NLSy1s) and 0.59 (RQ-BLSy1s), with slopes of 0.84 and 0.67, respectively (Table~\ref{tab:coef}). The marginal difference in $\CGalDust$ between the subclasses likely reflects variations in host galaxy properties. RL-BLSy1s occupy the same parameter space as RQ-BLSy1s, suggesting that jet emission does not dominate the MIR colors.

\subsection{Black hole mass, Bolometric Luminosity, and Eddington ratio}\label{sec:agn_params}
\begin{figure*}[hbt!]
\centering
\plotone{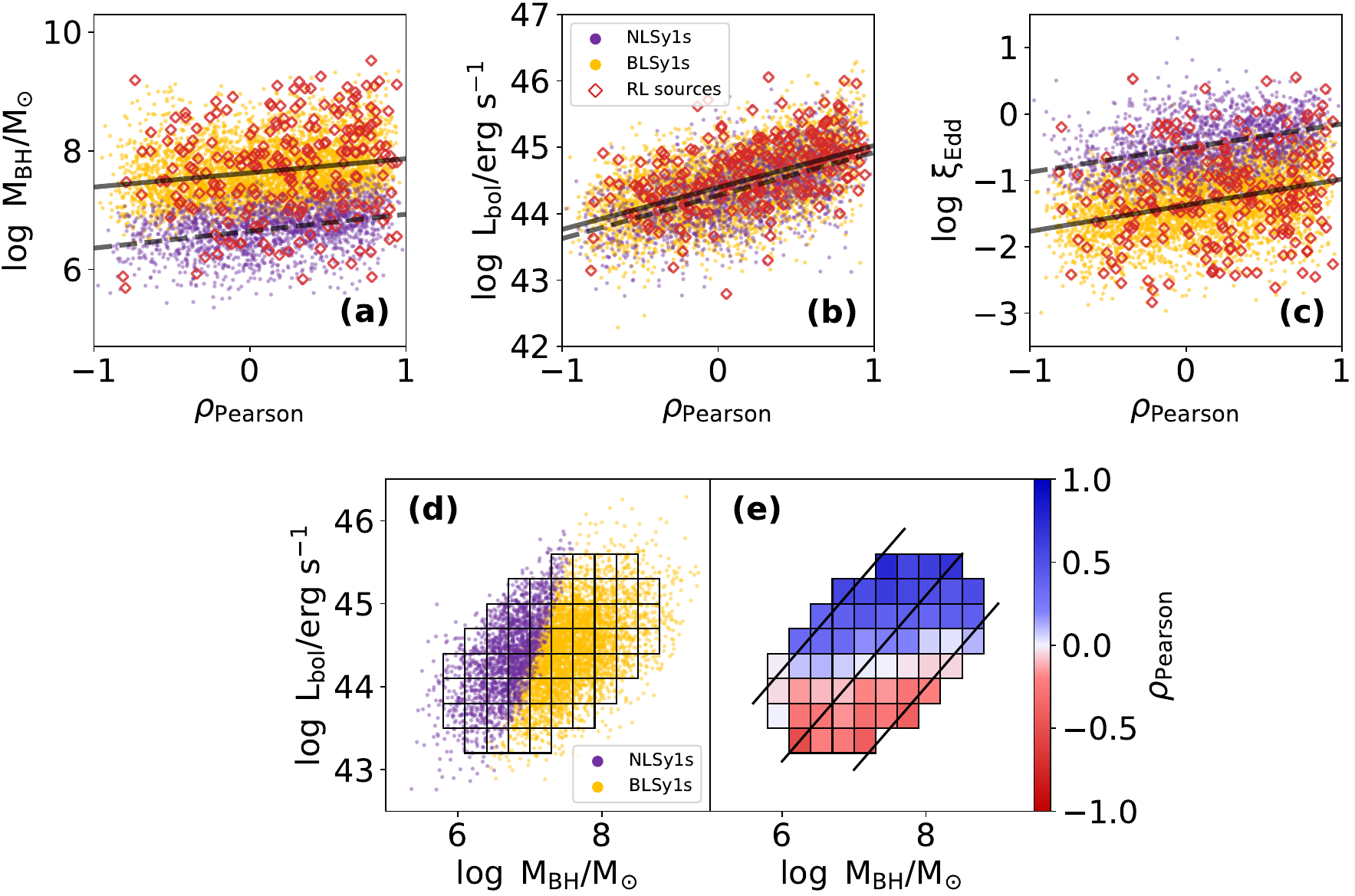}
\caption{
Correlations between the MCV coefficient $\coef$ and fundamental AGN parameters for the Seyfert~1 sample. 
(a--c) $\coef$ as a function of black hole mass ($\MBH$), bolometric luminosity ($\Lbol$), and Eddington ratio ($\REdd$). Symbol and color definitions are the same as in Figure~\ref{fig:mcv_analy}. The linear regression results for the RQ population are shown by dashed (NLSy1) and solid (BLSy1) lines.
(d) Distribution of radio-quiet Seyfert~1 (RQ-Sy1) galaxies in the $\MBH$--$\Lbol$ plane. The grid represents bins of 0.3~dex $\times$ 0.3~dex, with only bins containing $\geq 20$ sources shown. 
(e) Median $\coef$ calculated within each bin, with color-coding indicating the dominant variability trend (BWB vs. RWB). Diagonal lines represent loci of constant $\REdd$, with values increasing from the bottom-right to the top-left.
\label{fig:agn_param}}
\end{figure*}

In NLSy1s, $\Lbol$ was identified as the primary driver of MCV, while the correlations with $\MBH$ and $\REdd$ were likely driven by their coupling with $\Lbol$ (\paperI).
Figure~\ref{fig:agn_param} tests these trends for the BLSy1 population.
For BLSy1s, $\coef$ exhibits significant positive correlations with $\log \MBH$ ($r = 0.23$), $\log \REdd$ ($r = 0.37$), and $\log \Lbol$ ($r = 0.56$).
The correlation strengths with $\log \Lbol$ and $\log \REdd$ are comparable to those found for NLSy1s ($r = 0.55$ and $r = 0.40$, respectively), whereas the correlation with $\log \MBH$ is slightly weaker ($r = 0.33$). Linear fits yield comparable slopes for both subclasses ($\sim 0.26$ for $\log \MBH$, $\sim 0.64$ for $\log \Lbol$, and $\sim 0.38$ for $\log \REdd$).

To decouple the effects of these parameters, we performed a two-dimensional binning analysis on the $\MBH$--$\Lbol$ plane (Figure~\ref{fig:agn_param}d--e). The results confirm that: (1) at fixed $\Lbol$, $\coef$ shows no significant trend with $\MBH$; (2) $\coef$ increases systematically with $\Lbol$; and (3) lines of constant $\log (\Lbol/\MBH)$ show progressively stronger BWB trends, confirming that $\REdd$ modulates the MCV behavior through its coupling with luminosity.

\begin{figure*}[hbt!]
\centering
\includegraphics[width=1.6\columnwidth]{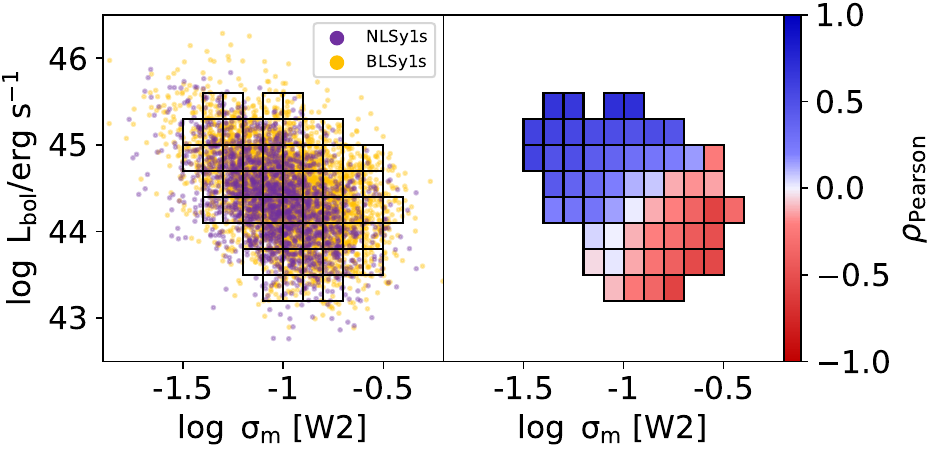}
\caption{
Dependence of MCV behavior on intrinsic variability and bolometric luminosity.
Left: Distribution of individual RQ-Sy1s in the $\Lbol$--$\sigma_m[W2]$ parameter space. Rectangular bins indicate the 0.3~dex $\times$ 0.3~dex regions used for ensemble analysis. 
Right: Median $\coef$ for each bin, demonstrating that the negative correlation between $\sigma_m[W2]$ and $\coef$ persists even after controlling for $\Lbol$.
\label{fig:mcv_sigma_m}}
\end{figure*}

The relationship between MIR variability and AGN properties is inherently complex \citep[e.g.,][]{Li2018,Son2023}. A potential explanation for the negative correlation between the $W2$ intrinsic variability amplitude ($\sigma_m[W2]$) and $\coef$ (Section~\ref{sec:sigma_m}) is that both parameters are driven by the bolometric luminosity, which we have identified as the primary driver of MCV. 
To test this hypothesis, we performed a two-dimensional binning analysis of the radio-quiet Seyfert~1 sample in the $\sigma_m[W2]$--$\Lbol$ plane, using a bin size of $0.3 \times 0.3$~dex. As shown in Figure~\ref{fig:mcv_sigma_m}, the negative correlation between $\sigma_m[W2]$ and $\coef$ persists even at a fixed $\Lbol$, indicating that luminosity is not the sole driver of this relationship.

\subsection{Distribution of MIR Color Variation Distribution}\label{sec:cplane}

We analyze the minimum MIR color ($C_{\min}$) and the color variation range ($\Delta C = C_{\max} - C_{\min}$). Previous work (\paperI) showed that host-galaxy starlight significantly affects RWB sources. These sources exhibit bluer $C_{\min}$ and larger $\Delta C$ compared to BWB sources.

\begin{figure*}[hbt!]
\plotone{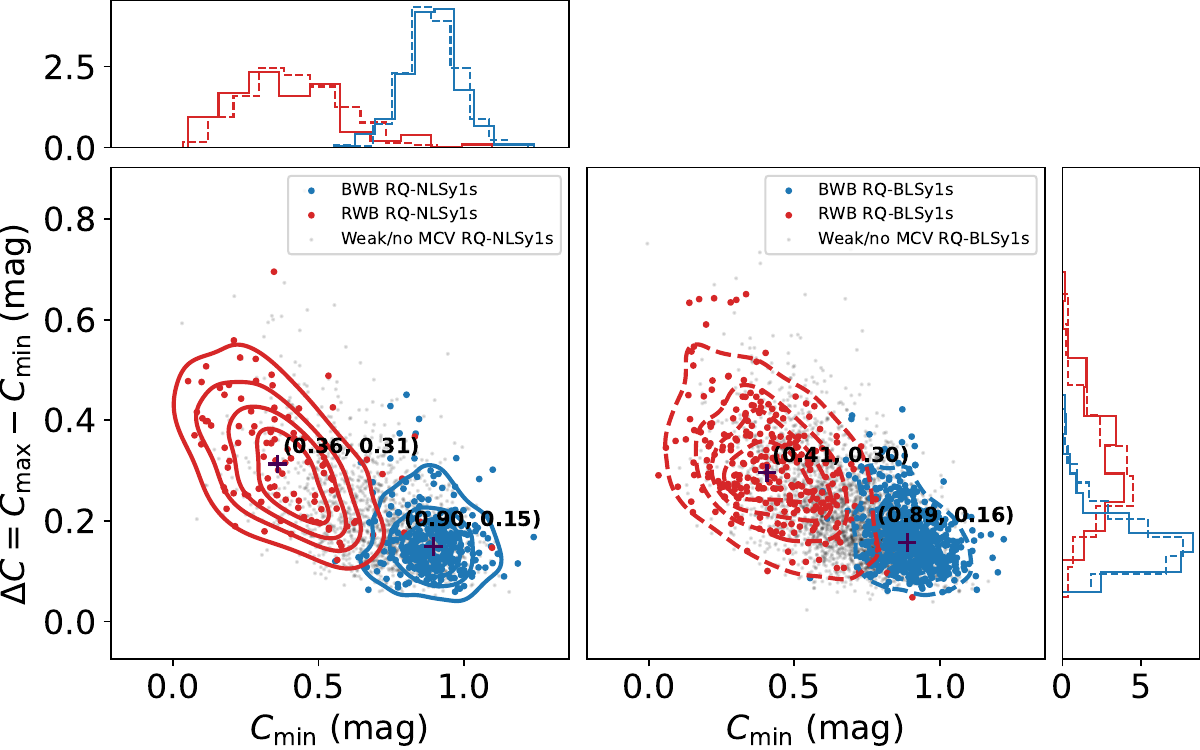}
\caption{
Distribution of the minimum MIR color ($C_{\min}$) versus the maximum color variation ($\Delta C = C_{\max} - C_{\min}$) for the Seyfert~1 sample. The populations are categorized into RWB RQ-Sy1s (red), BWB RQ-Sy1s (blue), and weak/no MCV RQ-Sy1s (gray). The distinct clustering of BWB sources at redder colors and RWB sources at larger variation ranges highlights the impact of host-galaxy contamination on the observed MIR color trends.
\label{fig:cplane}}
\end{figure*}

Figure~\ref{fig:cplane} shows that BLSy1s and NLSy1s share similar MCV distributions. The RWB radio-quiet population has a broad distribution. It is centered at $C_{\min} = 0.39$ and $\Delta C = 0.31$. In contrast, BWB sources form a tight cluster. Their median values are $C_{\min} = 0.90$ and $\Delta C = 0.16$. 

These results confirm two key patterns. First, BWB sources consistently show redder MIR colors ($C_{\min} \gtrsim 0.7$). Second, RWB sources show larger and more dispersed $\Delta C$ values. Weak or no MCV sources occupy an intermediate region, with median values of $C_{\min} = 0.71$ and $\Delta C = 0.19$. They are mostly located near the BWB population.

\section{The Fraction of RWB Sources}\label{sec:mcv_fraction}

Analysis in Section~\ref{sec:mcv_correlation} indicates that $\Lbol$ is the primary parameter driving MCV behavior in both NLSy1s and BLSy1s. This result is consistent with the findings of \paperI. However, $\Lbol$ alone cannot fully account for the observed MCV properties. The overall RWB fraction differs significantly between the two subclasses: 5.3\% for NLSy1s and 6.8\% for BLSy1s (Fisher exact test $p= 0.0015$). As shown in Figure~\ref{fig:agn_param}b, the $\Lbol$ distributions for both subclasses are similar, with a mean $\log \Lbol \approx 44.4$ and a standard deviation of $\approx 0.5$. This suggests that the difference in RWB fractions is not due to luminosity. Instead, the systematic offset in $\REdd$ suggests a secondary dependence on the accretion state.

\begin{figure*}[hbt!]
\centering
\includegraphics[width=1.6\columnwidth]{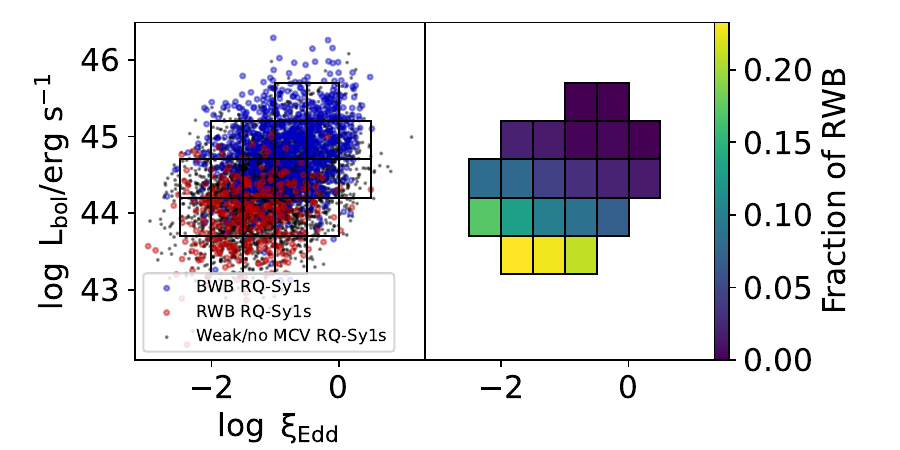}
\caption{
Analysis of the RWB fraction in the $\REdd$--$\Lbol$ plane.
Left: Distribution of individual RQ-Sy1s. The grid represents bins of 0.5~dex $\times$ 0.5~dex containing $\geq 70$ sources.
Right: The RWB fraction ($f_{\mathrm{RWB}}$) within each bin. The color scale indicates the fraction value. At a fixed luminosity, the RWB fraction decreases as the Eddington ratio increases.
\label{fig:frac_rwb}
}
\end{figure*}

To isolate the role of $\REdd$ from $\Lbol$, we performed a two-dimensional binning analysis in the $\Lbol$--$\REdd$ plane for the radio-quiet sample. We adopted a bin size of $0.5 \times 0.5$~dex and required at least 70 sources per bin to reduce random errors. For each bin, we calculated the RWB fraction, $f_{\mathrm{RWB}} = N_{\mathrm{RWB}} / N_{\mathrm{total}}$. Figure~\ref{fig:frac_rwb} (left) shows the source distribution in this plane, while Figure~\ref{fig:frac_rwb} (right) displays the resulting $f_{\mathrm{RWB}}$ values.

A clear trend emerges from this analysis. At a fixed $\REdd$, $f_{\mathrm{RWB}}$ decreases as $\Lbol$ increases. This is a natural consequence of $\Lbol$ driving the MCV: higher luminosities favor BWB trends, which suppresses the incidence of RWB sources. More importantly, at a fixed $\Lbol$, $f_{\mathrm{RWB}}$ also decreases systematically with increasing $\REdd$. This indicates that $\Lbol$ is not the only parameter determining MCV behavior. These results explain the different RWB fractions in NLSy1s and BLSy1s. Because NLSy1s are typically high-$\REdd$ systems, they fall into bins with lower $f_{\mathrm{RWB}}$. Conversely, the lower $\REdd$ of BLSy1s leads to a higher overall RWB fraction.

Partial correlation analysis further supports these findings. We calculated the partial correlation between $\coef$ and $\log \Lbol$ while controlling for $\REdd$. The correlation remains strong ($r = 0.496$, p-value $\ll 0.001$), confirming $\Lbol$ as the primary driver. We also tested the partial correlation between $\coef$ and $\log \REdd$ while controlling for $\Lbol$. This correlation is weaker but still highly significant ($r = 0.134$, p-value $\ll 0.001$). This statistical significance confirms that the Eddington ratio influences MCV properties independently of luminosity.

\section{Ensemble Structure Function Analysis}\label{sec:esf}

The MIR variability of AGN results from the dust torus responding to fluctuations in the accretion disk. Observed variability thus depends on both the intrinsic disk variability and the geometric properties of the torus. As demonstrated by \citet{Li_and_Shen2023}, this relationship allows for statistical inference of the typical torus size.

The structure function, $\mathrm{SF}(\Delta t)$, quantifies the variability amplitude as a function of the rest-frame time lag, $\Delta t$. By calculating an ensemble SF from a large sample, we obtain a measure of variability linked to the physical scale of the torus. The torus acts as a low-pass filter, smoothing the high-frequency optical/UV variations from the central engine. A larger torus inner radius results in stronger smoothing, which leads to a steeper MIR structure function \citep{Li_and_Shen2023}.
Reverberation mapping (RM) studies show that higher $\Lbol$ increases the dust torus radius, while many observations confirm a corresponding steepening of the MIR SF \citep[e.g.,][]{Son2023,Kim2024}.

Recent work indicates that NLSy1s exhibit flatter MIR ensemble SFs compared to BLSy1s \citep{Hu_and_Mao2025}. Since the $\Lbol$ distributions of the two subclasses in our sample are similar (Section~\ref{sec:mcv_fraction}), this difference cannot be attributed to luminosity. Motivated by the distinct $\REdd$ distributions noted previously, we applied a two-dimensional binning analysis (bin size = 0.5~dex $\times$ 0.5~dex, $>70$ sources per bin) to the $\Lbol$--$\REdd$ plane. We define the ensemble structure function as:
\begin{equation}
{\rm SF}^2(\Delta t) = \frac{1}{N_{\Delta t}} \sum^{N_{\Delta t}} \left[ (m(t+\Delta t) - m(t))^2 - (\varepsilon^2_e(t+\Delta t) + \varepsilon^2_e(t)) \right],
\end{equation}
where $m(t)$ and $\varepsilon_e(t)$ represent binned magnitudes and uncertainties at rest-frame time $t$.

\begin{figure*}[hbt!]
\plotone{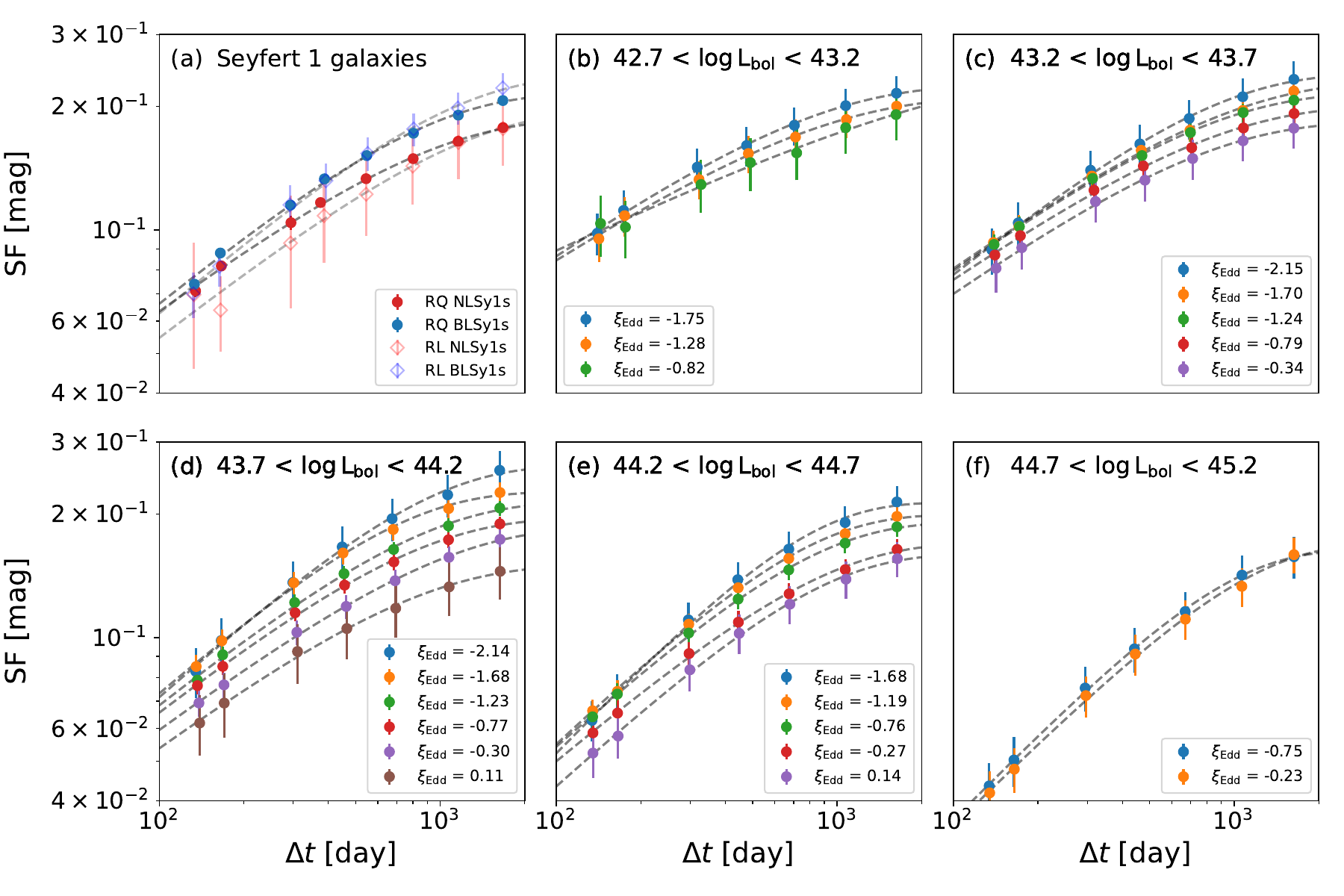}
\caption{Ensemble structure functions for the Seyfert~1 sample. 
(a) SFs for RQ and RL subsamples. Dashed lines represent the best-fit models. 
(b--f) SFs for subsamples binned in the $\Lbol$--$\REdd$ plane. The flattening of the slope with increasing Eddington ratio is evident across luminosity bins.
\label{fig:esf}
}
\end{figure*}

We first computed the ensemble SFs for RQ NLSy1s and BLSy1s. Figure~\ref{fig:esf}a shows that both exhibit a power-law rise at short lags and a plateau at longer lags. We modeled the data using the functional form:
\begin{equation}
\mathrm{SF}(\Delta t) = \mathrm{SF}_{\infty}~\left[1 - e^{-(\Delta t / \tau)^\beta}\right],
\end{equation}
where $\beta$ characterizes the slope of the power-law portion. We find $\beta = 1.07 \pm 0.03$ for RQ-NLSy1s and $\beta = 1.18 \pm 0.01$ for RQ-BLSy1s. The NLSy1 SF is significantly flatter and has a lower amplitude ($\mathrm{SF}_{\infty} = 0.19 \pm 0.02$ mag) compared to BLSy1s ($\mathrm{SF}_{\infty} = 0.22 \pm 0.01$ mag). These results are consistent with prior studies and align with the lower optical/UV variability observed in NLSy1s \citep{Ai2013, Rakshit2017}. The SFs of RL counterparts are statistically indistinguishable from the RQ samples when $1\sigma$ uncertainties are considered.

\begin{figure}[hbt!]
\plotone{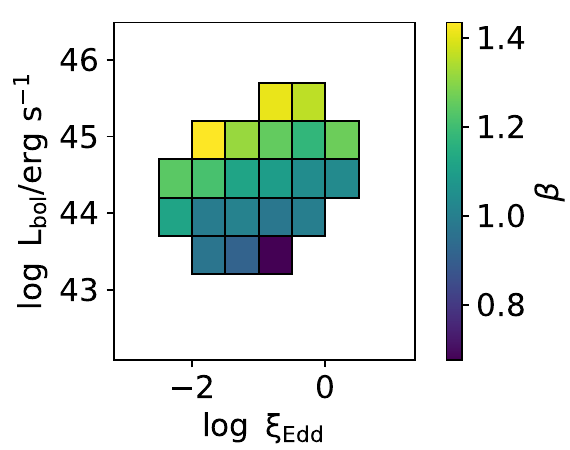}
\caption{
Structure function slope ($\beta$) binned in the $\Lbol$--$\REdd$ plane. 
Each bin contains at least 70 sources. Two distinct trends are visible: 
(1) at fixed $\REdd$, $\beta$ increases with $\Lbol$; 
(2) at fixed $\Lbol$, $\beta$ decreases with increasing $\REdd$. 
This confirms that the Eddington ratio independently modulates the dust torus response.
\label{fig:esf_beta}
}
\end{figure}

Next, we fitted the ensemble SF for each $\Lbol$--$\REdd$ bin (Figures~\ref{fig:esf}b--f). The derived $\beta$ values, shown in Figure~\ref{fig:esf_beta}, demonstrate that at a fixed $\Lbol$, the SF becomes flatter as $\REdd$ increases. This confirms that the accretion state modulates variability properties independently of luminosity. 

We note that \citet{Hu_and_Mao2025} reported an opposite trend within their subsamples, where the SF slope steepened with increasing $\REdd$. However, they did not control for $\Lbol$. Because both $\beta$ and $\REdd$ correlate positively with $\Lbol$, the dominant $\Lbol$--$\beta$ correlation can mask the intrinsic negative $\REdd$--$\beta$ correlation. By holding $\Lbol$ fixed, our binning analysis reveals the true anti-correlation between $\beta$ and $\REdd$.

The ensemble structure function analysis, combined with the results in Section~\ref{sec:mcv_fraction}, consistently demonstrates that bolometric luminosity is the primary driver of the observed MIR color variations and structure function properties. However, the Eddington ratio clearly plays a secondary but significant modulating role. Specifically, the accretion state independently influences both the incidence of color trends and the steepness of the variability power spectrum. A detailed discussion of the physical implications, including the inferred geometry and scale of the dust torus, is presented in Section~\ref{sec:scale_dt}.

\section{Discussion}\label{sec:discussion}

\subsection{Mid-infrared Color Variation in Seyfert 1 Galaxies}

\paperI{} proposed that MIR emission in NLSy1s arises from two components: AGN-heated dust and invariant host-galaxy starlight ($F_{\lambda,\mathrm{invar}}$). This combination explains both BWB and RWB trends. When the AGN is active, red thermal emission from hot dust dominates, leading to a BWB trend. When nuclear activity is low, blue host starlight becomes significant. A subsequent increase in luminosity heats the dust, shifting the color from ``blue'' starlight to ``red'' hot dust, which manifests as an RWB trend.

Our results show that BLSy1s exhibit MCV correlations nearly identical to those of NLSy1s (Section~\ref{sec:mcv_correlation} and Table~\ref{tab:coef}). For both subclasses, $\Lbol$ is the primary parameter governing MCV behavior (Figure~\ref{fig:agn_param}e). This confirms that host-contaminated dust reprocessing is a universal mechanism across Seyfert 1 galaxies.

However, $\Lbol$ is not the only factor. After controlling for luminosity, $\REdd$ correlates with both the RWB fraction and the MIR SF slope (Sections~\ref{sec:mcv_fraction} and \ref{sec:esf}). Higher-$\REdd$ systems show lower RWB fractions and flatter SFs. This indicates that the accretion rate modulates dust torus properties in ways that luminosity alone cannot. We discuss this further in Section~\ref{sec:scale_dt}.

Additionally, we find no significant difference in MCV between radio-loud and radio-quiet sources. Although jets contribute to MIR emission in RL AGNs, their variability usually occurs on short timescales \citep[e.g.,][]{Jiang2012}. Over the multi-year WISE baseline, these fluctuations are likely smoothed out, leaving the thermal dust reverberation as the dominant signal.

\subsection{The $\sigma_{m}[W2]$--$\coef$ Correlation}
We find a negative correlation between the $W2$ variability amplitude ($\sigma_m[W2]$) and the MCV coefficient $\coef$ (Figure~\ref{fig:mcv_analy}c--d).
The MCV trend naturally favors such an anti-correlation: for a given $W1$ variability amplitude, RWB behavior (lower $\coef$) is associated with a larger $W2$ variation relative to $W1$, whereas BWB behavior (higher $\coef$) corresponds to a smaller one. Thus, even at fixed $\Lbol$, an anti-correlation between $\sigma_m[W2]$ and $\coef$ can arise as long as sources exhibit different MCV trends (Figure~\ref{fig:mcv_sigma_m}), which may result from variations in other properties, such as $\REdd$ and the host-galaxy fraction.

\begin{figure*}[hbt!]
\plotone{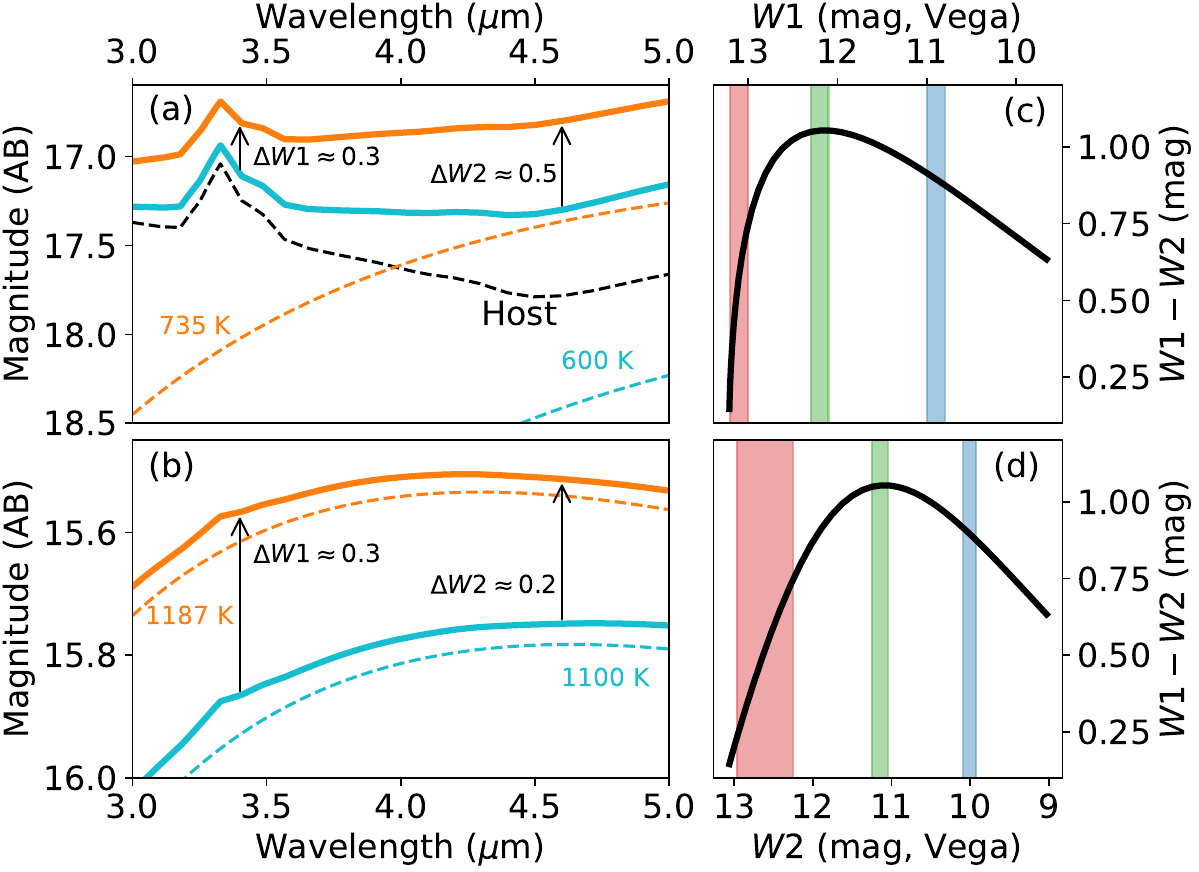}
\caption{
Schematic illustration of the host-contaminated dust reprocessing model presented in \paperI{}. Panels (a) and (b) show the SED evolution for cases with higher and lower host contributions, respectively, with the $y$-axis in AB magnitudes. The model parameters are identical to those used in Figure~6 of \paperI{}. The black dashed lines represent the host from the spiral-galaxy template of \citet{Assef2010}, while the cyan and orange dashed lines denote the low- and high-temperature blackbody components; the corresponding solid lines show the total flux. The temperature variations yield $\Delta W1\sim0.3$ mag in both cases, while $\Delta W2=0.5$ and $0.2$ mag for panels (a) and (b), respectively, indicating RWB ($\Delta W2>\Delta W1$) and BWB ($\Delta W2<\Delta W1$) behavior. Panels (c) and (d) show the color-magnitude relation generated by the \paperI{} model, with $W1-W2$ plotted against $W1$ and $W2$, respectively. Panel (c) shows the RWB (red), weak/no MCV (green), and BWB (blue) regimes, with shaded vertical bands indicating an identical $W1$ variability amplitude; panel (d) shows the corresponding $W2$ responses. The steep RWB relation allows sources with different MCV trends to exhibit more distinct $W2$ responses, thereby producing a stronger anti-correlation between $\sigma_m[W2]$ and $\coef$.
\label{fig:model_sigma_m}
}
\end{figure*}

As demonstrated by the host-contaminated dust reprocessing model in \paperI{}, in which the AGN-heated dust is represented by a single blackbody and the host by the spiral-galaxy template of \cite{Assef2010}, varying the relative host contribution can produce RWB or BWB behavior. Following the same model and parameter settings adopted in \paperI{} (see their Section~5.1 and Figure~6), we show the corresponding SED evolution in Figure~\ref{fig:model_sigma_m}.
Panel~(a) shows an RWB case with a high host contribution ($\sim67\%$--$88\%$ in the $W1$ band) and a blackbody temperature increase from $600$ to $735$~K, while panel~(b) shows a BWB case with a lower host contribution ($\sim4\%$--$6\%$ in the $W1$ band) and a temperature increase from $1100$ to $1187$~K.
In both cases, the temperature variations are chosen to produce a similar $W1$ variability amplitude ($\sim0.3$~mag). The corresponding $W2$ variability amplitude is substantially larger in the RWB case ($\sim0.5$~mag) than in the BWB case ($\sim0.2$~mag), as expected from their respective MCV trends.

In addition, as illustrated by the color-magnitude relations produced by the \paperI{} model (also shown in Figure~\ref{fig:model_sigma_m}c--d), the relatively steep relation in the RWB regime allows sources with different $\coef$ to exhibit substantially different $W2$ variability amplitudes, helping produce a sufficiently strong anti-correlation between $\sigma_m[W2]$ and $\coef$.

\subsection{Dust Grain Properties and Torus Scaling}
In \paperI{}, we modeled hot dust as an ideal blackbody. To account for grain composition and size --- factors that significantly impact the MIR spectral shape --- we here adopt a modified blackbody model:
\begin{equation}
F_{\lambda}(T_d) = F_{\lambda,\mathrm{invar}} + A \cdot Q_{\text{abs}}(\lambda; a) B_\lambda(T_d).
\end{equation}
Here, $T_d$ is the dust temperature, $a$ is the grain radius, and $Q_{\text{abs}}$ is the absorption efficiency. We consider graphite, silicates, and silicon carbide with radii of 0.01, 0.1, and 1.0~$\mu$m, using efficiencies from \citet{1993ApJ...402..441L}. The invariant component $F_{\lambda,\mathrm{invar}}$ is calibrated to represent a 50\% starlight fraction at $W1$ for the median source in our sample. 

\begin{figure}[hbt!]
\plotone{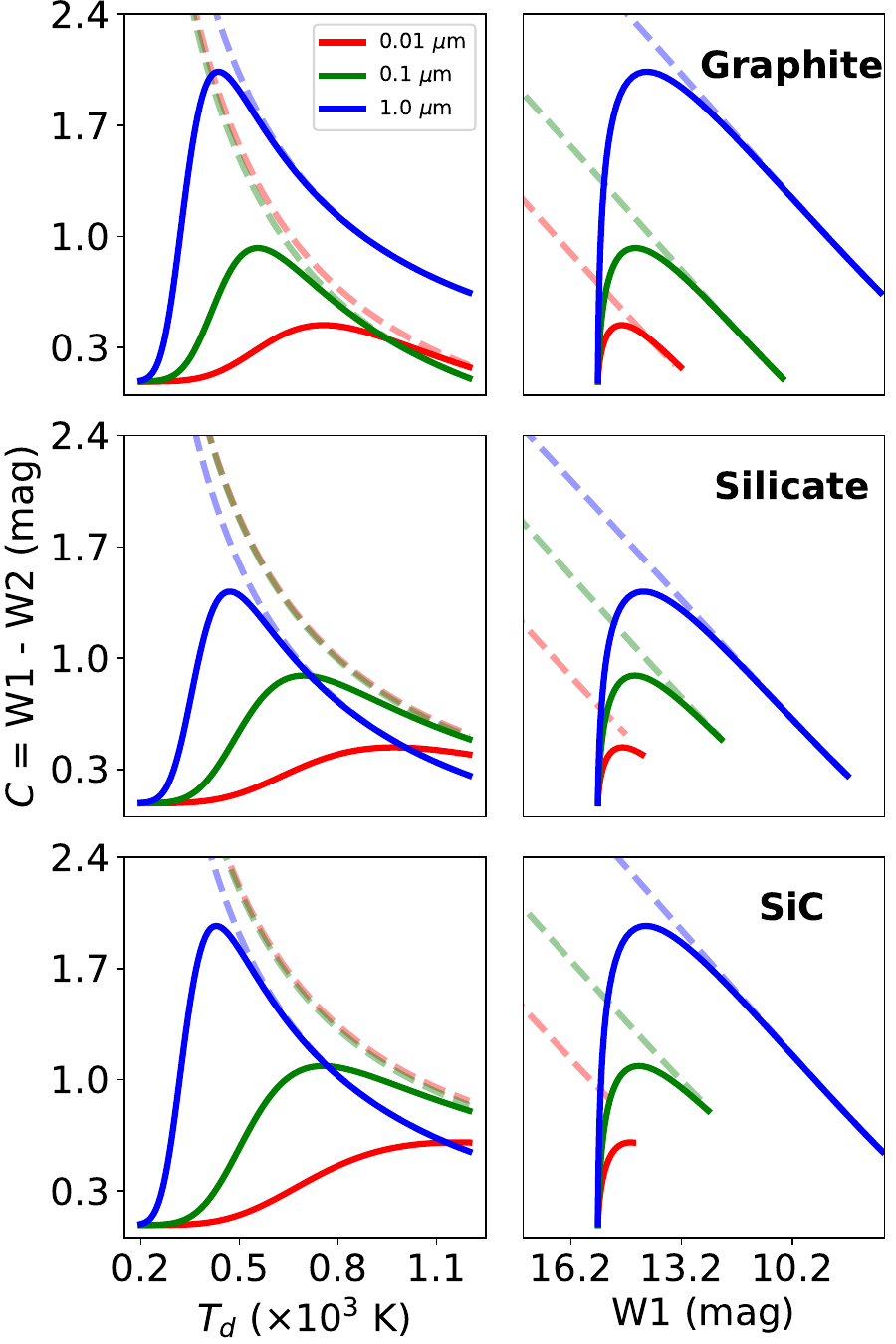}
\caption{
\label{fig:model}
Modeled MIR color properties for various dust compositions and grain radii. \textit{Left column}: MIR color ($C = W1 - W2$) as a function of dust temperature ($T_d$) for the modified blackbody model. Each row represents a specific dust composition: graphite (top), silicates (middle), and silicon carbide (bottom). Colored lines denote different grain radii ($a$), as indicated in the legend. The dashed black line in each panel represents the intrinsic dust color without host starlight contamination ($F_{\rm invar} = 0$). \textit{Right column}: Corresponding color-magnitude diagrams showing $C$ as a function of $W1$ magnitude, calculated under the same model assumptions.
}
\end{figure}

Our modeling (Figure~\ref{fig:model}) indicates that larger grains produce redder MIR colors at a given $W1$ magnitude. The smallest grains ($a = 0.01~\mu$m) fail to reproduce the observed median colors ($C_{\rm min} \approx 0.9$) and the large color variations ($\Delta C$) seen in RWB sources. We conclude that the grains responsible for the observed MCV must have sizes $a \gtrsim 0.1~\mu$m.

This size limit is an order of magnitude larger than the mean radius of the interstellar MRN distribution ($\sim 0.008~\mu$m; \citealt{MRN1977}). This discrepancy is physically consistent with the harsh AGN environment. In intense radiation fields, small grains are preferentially destroyed by sublimation, while larger grains survive closer to the central engine due to lower equilibrium temperatures. Recent \textit{JWST} observations of NGC~7469 \citep{Lai2022} support this, showing an increasing fraction of large, ionized grains toward the nucleus. 

While our single-component model is a simplification, it provides a valuable lower limit on grain size. Combining this absolute limit with empirical grain-size ratios from variability analyses \citep[e.g.,][]{Long2026} may further constrain the complex mixture of grain species in future torus models.

For BWB sources, the MIR emission is dominated by AGN-heated dust. We therefore approximate the radiation as originating solely from hot dust and apply the modified blackbody model to estimate the dust temperature ($T_d$) and total grain number ($N_d$). To minimize host-galaxy contamination, we utilize only the brightest $W1$ epoch and its corresponding $W2$ measurement for each source.

\begin{figure}[hbt!]
\centering
\plotone{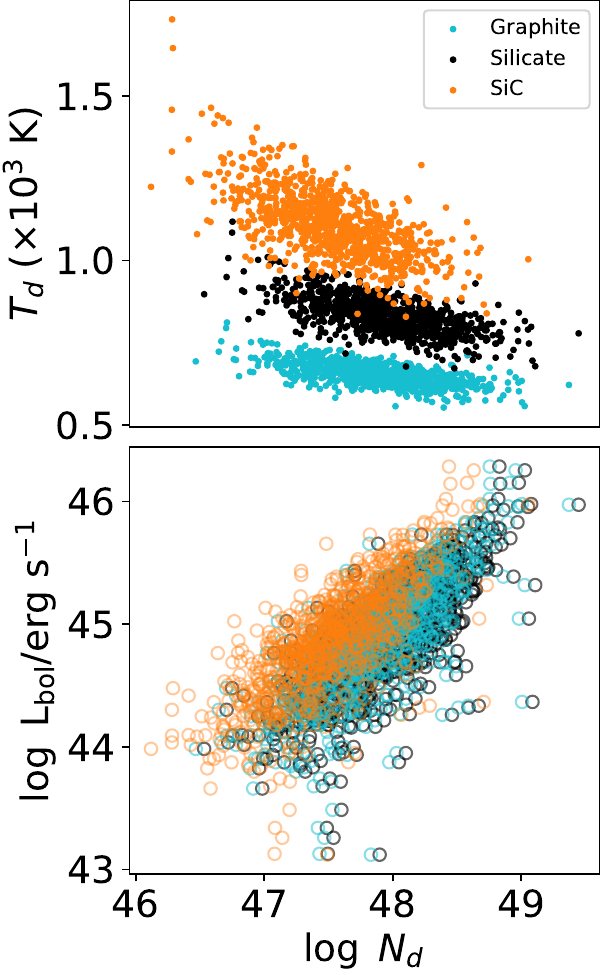}
\caption{
Results of the modified blackbody fitting for the BWB radio-quiet sample. \textit{Top}: Derived dust grain number ($N_d$) as a function of dust temperature ($T_d$). \textit{Bottom}: Relationship between bolometric luminosity ($L_{\rm bol}$) and $N_d$. In both panels, colors indicate the assumed dust composition: graphite (cyan), silicates (black), and silicon carbide (orange).
\label{fig:mblackbody}}
\end{figure}

The top panel of Figure~\ref{fig:mblackbody} displays the distribution of $\log N_d$ versus $T_d$ across the three grain compositions. For all species, the derived temperatures generally remain below their respective sublimation limits. We find a significant negative correlation between $\log N_d$ and $T_d$, with correlation coefficients of $r \approx -0.65$ for graphite, silicates, and SiC. Physically, this anti-correlation suggests that dust situated further from the central engine is cooler. While the dust density typically decreases with distance, the expanding volume of the torus results in a larger total grain count. Consequently, $N_d$ likely scales with the inner radius ($R_{\rm in}$) of the dust torus, making the observed $N_d$--$T_d$ relation a proxy for the underlying $R_{\rm in}$--$T_d$ connection.

Furthermore, we find a significant positive correlation between $\log N_d$ and the bolometric luminosity $L_{\rm bol}$ (Figure~\ref{fig:mblackbody}, bottom). The linear regression yields:
\begin{equation}
\log N_d = a\cdot\log \left( \frac{L_{\rm bol}}{\rm erg~s^{-1}} \right) + b,
\end{equation}
where $a \approx 0.72$ for all three compositions ($a = 0.72 \pm 0.02$ for graphite/silicates and $0.73 \pm 0.02$ for SiC). Given that $N_d$ is proportional to $R_{\rm in}$, this correlation reflects the established positive relationship between the geometric scale of the dust torus and the AGN luminosity.

Recent reverberation mapping measurements have established a size--luminosity relation of the form $\log R \propto 0.5 \log \Lbol$ \citep[e.g.,][]{Koshida2014, Lyu2019, Yang2020,Tomar2025}. Similar results have been obtained through statistical analyses of MIR ensemble structure functions \citep{Li_and_Shen2023, Kim2024}. Assuming the dust grain number density $n_d$ follows a radial power-law distribution, $n_d = n_0 (R/R_0)^{-p}$ for $R_{\rm in} < R < R_{\rm out}$, the total number of dust grains is given by:
\begin{equation}
N_d = \frac{4\pi\sigma}{90^{\circ}} n_0 R_0^3 \frac{Y^{3-p} - 1}{3-p} \left( \frac{R_{\rm in}}{R_0} \right)^{3-p},
\end{equation}
where $Y = R_{\rm out}/R_{\rm in}$ and $\sigma$ is the covering factor. 

If the variation in $N_d$ across our sample is driven primarily by changes in $R_{\rm in}$ (rather than $n_0$, $\sigma$, or $Y$), we obtain the following proportionality: 
\begin{equation}
\log R_{\rm in} \propto \frac{1}{3-p} \log N_d \propto \frac{a}{3-p} \log L_{\rm bol}.
\end{equation}
By combining our derived value of $a \approx 0.72$ with the size--luminosity relation slope of 0.5, we find a density power-law index of $p \approx 1.56$.

Our result is consistent with MIR interferometric constraints from \citet{Kishimoto2011}. They found radial surface density profiles between $\Sigma(r) \propto r^{-1}$ and $\Sigma(r) \propto r^{0}$ for Type~1 AGNs. For a linearly flaring torus, $\Sigma(r) \propto r^{1-p}$ corresponds to $p \sim 1$--$2$. Our independent estimate of $p$ falls within this range, supporting the physical assumptions of our model.

\subsection{Linking Torus Scale to Accretion Rate}\label{sec:scale_dt}

Our analysis of MCV trends (Sections~\ref{sec:agn_params} and \ref{sec:mcv_fraction}) and structure functions (Section~\ref{sec:esf}) demonstrates that while $\Lbol$ is the primary driver of MIR behavior, it is not the sole controlling parameter. Within fixed $L_{\rm bol}$ bins, both the RWB fraction and the SF slope $\beta$ correlate negatively with $\REdd$ (see Figures~\ref{fig:frac_rwb} and \ref{fig:esf_beta}). 

This dependency suggests that the physical scale of the dust torus varies with the accretion rate. In high-$\REdd$ systems, such as NLSy1s, the circumnuclear dust torus may be situated closer to the central black hole. For a given $\Lbol$, a smaller torus radius leads to higher equilibrium dust temperatures. This proximity favors a BWB response to luminosity fluctuations, as the hot dust spectral energy distribution (SED) peak shifts blueward upon heating.

This interpretation is supported by variability modeling. \citet{Li_and_Shen2023} demonstrated that the MIR SF slope is primarily governed by the inner radius of the dust torus, where a smaller radius produces a flatter SF. Our observation of flatter SFs in high-$\REdd$ systems is consistent with this model.

The higher accretion rates of these systems may also affect the torus geometry through radiation pressure. Although stronger radiation pressure at higher Eddington ratios can drive dusty outflows, dust near the equatorial plane is less directly exposed to the central radiation field and may therefore be less susceptible to outward displacement. Radiation pressure may instead preferentially affect dust at higher elevations, primarily altering the covering factor rather than significantly increasing the inner dust radius. As introduced in Section~\ref{sec:intro}, previous studies have found that the torus covering factor decreases with increasing Eddington ratio, likely due to radiation-driven outflows that clear circumnuclear dust (e.g., \citealt{Ricci2017,Ricci2023,Mizukoshi2024,Zhuang2018}).

Because the $\Lbol$ distributions of our NLSy1 and BLSy1 samples are comparable, the observed differences in RWB fractions and SF slopes cannot be attributed to luminosity. Instead, these differences appear to be driven by the distinct Eddington ratios of the two subclasses, implying a systematic difference in their torus scales. This result may hint at an evolutionary sequence: NLSy1s accreting at high Eddington ratios gradually consume their circumnuclear fuel and grow their black hole masses. As the inner torus is depleted and the reservoir recedes, the system eventually evolves into a lower-rate BLSy1.

\subsection{Changing-look AGNs}

In \paperI, changing-look AGNs (CLAGNs) were found to occupy a region of the $C_{\rm min}$--$\Delta C$ plane that overlaps significantly with RWB NLSy1s. We here extend this finding to the broader Seyfert~1 population. Both ``turn-off'' and ``turn-on'' CLAGNs undergo low-accretion states where the MIR emission is dominated by host-galaxy starlight. Consequently, their long-term MCV behavior mirrors the RWB trend.

This connection suggests that MCV may provide a useful means of pre-selecting Seyfert~1s that have undergone substantial long-term changes in nuclear activity. Large MIR color excursions associated with extreme accretion-state transitions can enhance the detectability of MCV. However, the presence of an RWB trend itself is not necessarily a direct indicator of a changing-look transition, as ordinary AGN variability combined with different levels of host contribution can also produce RWB trends of varying strength.

If we conservatively assume that all Seyfert~1s with $\coef < 0$ could be associated with CLAGN-like long-term variability, up to $39.2\%$ of our sample would satisfy this broad MIR variability criterion. Adopting our definition of RWB Seyfert~1s, $\coef < -0.6$, which corresponds to a statistically significant RWB signal (Section~\ref{sec:mc_estimation}), this upper limit decreases to $6.3\%$. These fractions should not be interpreted as direct measurements of the intrinsic CLAGN incidence rate, but rather as upper limits on the fractions of Seyfert~1s that could be flagged as potential CLAGN candidates based on their MCV. Spectroscopic follow-up remains necessary for confirmation.

\section{Conclusions} \label{sec:conclusions}

We have presented a systematic comparative analysis of mid-infrared color variations (MCVs) in a sample of 1,977 NLSy1s and 4,113 BLSy1s. These sources were monitored over a 14-year baseline by WISE and NEOWISE. This study extends the framework established in \paperI\ to the broader Seyfert~1 population. Our principal conclusions are as follows:

\begin{enumerate}
\item \textit{Universality of the MCV Mechanism.} The MCV properties of BLSy1s are fundamentally consistent with those of NLSy1s. For both subclasses, bolometric luminosity ($\Lbol$) is the primary driver of MCV behavior. This confirms that host-contaminated dust reprocessing is a universal mechanism across Seyfert~1 galaxies. In this model, observed color variations are driven by temperature fluctuations in circumnuclear dust in response to variable accretion luminosity, with the color-magnitude relation modulated by invariant host-galaxy emission.

\item \textit{The Role of Accretion State.} While $\Lbol$ determines the overall sign and strength of the color-magnitude correlation, the relative frequency of BWB and RWB behaviors is modulated by the Eddington ratio ($\REdd$). At a fixed $\Lbol$, higher-$\REdd$ sources exhibit systematically lower RWB fractions. This indicates that the accretion state influences the physical conditions of the dust-emitting region beyond the effects of luminosity alone. We suggest that this dependence may be partly related to changes in the inner radius of the MIR-emitting dust distribution: a smaller inner scale at higher $\REdd$ would produce hotter dust and favor a BWB response. The role of radiation pressure may also be incorporated into this picture. Although stronger radiation pressure at higher $\REdd$ can drive dusty outflows, dust near the equatorial plane is less directly exposed to the central radiation field and may be less susceptible to outward displacement. Radiation pressure may instead preferentially affect dust at higher elevations, altering the torus covering factor rather than significantly increasing the inner dust radius.

\item \textit{Torus Scaling and Eddington Ratio.} Analysis of the MIR ensemble structure functions (SFs) provides independent evidence for torus scaling.
We confirm a positive correlation between the SF slope ($\beta$) and $\Lbol$, consistent with the expected increase in the dust sublimation radius with luminosity. Furthermore, we find a significant negative correlation between $\beta$ and $\REdd$ at fixed $\Lbol$. Higher-$\REdd$ sources therefore exhibit flatter MIR SFs, consistent with a smaller inner scale of the MIR-emitting dust distribution.

\item \textit{MIR Pre-selection of CLAGN Candidates.} CLAGNs exhibit MIR color-variation distributions similar to those of RWB sources. Assuming that this similarity reflects a common MCV behavior, we estimate that $\lesssim6.3\%$ of Seyfert~1s may be CLAGN candidates.
\end{enumerate}

In summary, our results establish thermal reprocessing as the universal mechanism for MIR variability in Seyfert~1 galaxies. We identify the Eddington ratio as a key secondary parameter that influences the physical scale and structure of the circumnuclear dust torus.

\begin{deluxetable}{cccc}
\tabletypesize{\small}
\tablecaption{Distribution of MCV Classifications}
\label{tab:number_mcv}
\tablehead{
    \colhead{(1)} & 
    \colhead{(2)} & 
    \colhead{(3)} & 
    \colhead{(4)} \\
    \colhead{MCV} &
    \colhead{RQ or RL} &
    \colhead{NLSy1 or BLSy1} &
    \colhead{Number}
}
\startdata
BWB & RQ & NLSy1 & 316 \\
BWB & RQ & BLSy1 & 614 \\
BWB & RL & NLSy1 & 14 \\
BWB & RL & BLSy1 & 45 \\
RWB & RQ & NLSy1 & 103 \\
RWB & RQ & BLSy1 & 268 \\
RWB & RL & NLSy1 & 2 \\
RWB & RL & BLSy1 & 6 \\
Weak/no MCV & RQ & NLSy1 & 1,508 \\
Weak/no MCV & RQ & BLSy1 & 3,055 \\
Weak/no MCV & RL & NLSy1 & 34 \\
Weak/no MCV & RL & BLSy1 & 125 \\
\enddata
\tablecomments{Distribution of the final sample (1,977 NLSy1s and 4,113 BLSy1s) categorized by mid-infrared color variability (MCV) class, radio-loudness (RQ/RL), and Seyfert~1 subclass.
}
\end{deluxetable}

\begin{deluxetable*}{ccccccccc}
\tablecaption{
Correlations between AGN Parameters and $\coef$}
\label{tab:coef}
\tablehead{
\colhead{Subsample} & \multicolumn{4}{c}{NLSy1s} & \multicolumn{4}{c}{BLSy1s} \\
\cline{2-5}\cline{6-9}
\colhead{} & \colhead{r} & \colhead{p-value} & \colhead{slope} & \colhead{intercept} & \colhead{r} & \colhead{p-value} & \colhead{slope} & \colhead{intercept}
}
\colnumbers
\startdata
{$\log\sigma_m[W1]$} &{-0.039}&{0.087}&{-0.017$\pm$0.010}&{-0.922$\pm$0.004}&{-0.135}&{$\ll 0.001$}&{-0.056$\pm$0.007}&{-0.857$\pm$0.003}\\
{$\log\sigma_m[W2]$} &{-0.472}&{$\ll 0.001$}&{-0.221$\pm$0.009}&{-0.885$\pm$0.004}&{-0.519}&{$\ll 0.001$}&{-0.244$\pm$0.006}&{-0.828$\pm$0.003}\\
{$\Cmean$} &{0.715}&{$\ll 0.001$}&{0.287$\pm$0.007}&{0.786$\pm$0.003}&{0.726}&{$\ll 0.001$}&{0.268$\pm$0.004}&{0.785$\pm$0.002}\\
{$\CGalDust$} &{0.679}&{$\ll 0.001$}&{0.845$\pm$0.023}&{3.997$\pm$0.010}&{0.588}&{$\ll 0.001$}&{0.667$\pm$0.016}&{4.056$\pm$0.007}\\
{$\log\MBH$} &{0.330}&{$\ll 0.001$}&{0.286$\pm$0.019}&{6.644$\pm$0.008}&{0.225}&{$\ll 0.001$}&{0.239$\pm$0.016}&{7.627$\pm$0.008}\\
{$\log\Lbol$} &{0.550}&{$\ll 0.001$}&{0.648$\pm$0.022}&{44.272$\pm$0.010}&{0.559}&{$\ll 0.001$}&{0.631$\pm$0.015}&{44.393$\pm$0.007}\\
{$\log\REdd$} &{0.400}&{$\ll 0.001$}&{0.363$\pm$0.019}&{-0.512$\pm$0.009}&{0.369}&{$\ll 0.001$}&{0.392$\pm$0.016}&{-1.373$\pm$0.007}\\
\enddata
\tablecomments{
Pearson correlation coefficients ($r$) and linear regression parameters (slope and intercept) between the MCV correlation coefficient $\coef$ and physical AGN properties for radio-quiet NLSy1s and BLSy1s.
}
\end{deluxetable*}

\begin{acknowledgments}
This work was supported by the National Science Foundation of China (NSFC-12573110, 12333002, 12233001, 12221003, 12133001 and 11833007) and the China Manned Space Program (CMS-CSST-2025-A09). YLA was supported by the Natural Science Foundation of Top Talent of SZTU (GDRC202208) and the Shenzhen Science and Technology program (JCYJ20230807113910021). LMD also acknowledges the support from the Key Laboratory for Astronomical Observation and Technology of Guangzhou, the Astronomy Science and Technology Research Laboratory of the education department of Guangdong Province. 
This research makes use of data products from the {\it Wide-field Infrared Survey Explorer (WISE)} and the {\it Near-Earth Object Wide-field Infrared Survey Explorer (NEOWISE)}. {\it WISE} is a joint project of the University of California, Los Angeles, and the Jet Propulsion Laboratory/California Institute of Technology; {\it NEOWISE} is a project of the Jet Propulsion Laboratory/California Institute of Technology. {\it WISE} and {\it NEOWISE} are funded by the National Aeronautics and Space Administration.

\end{acknowledgments}

\bibliography{sample701}{}
\bibliographystyle{aasjournalv7}
\end{document}